\documentclass[a4paper,11pt,oneside]{article}

\usepackage[english]{babel}
\usepackage[T1]{fontenc}
\usepackage[utf8]{inputenc}

\usepackage[pdftex,unicode]{hyperref}
\hypersetup{pdftitle=How Many Customers Does a Retail Store Have?}
\hypersetup{pdfauthor=Ondřej Sokol and Vladimír Holý}

\usepackage[margin=60pt]{geometry}
\usepackage{amsmath}
\usepackage{amssymb}
\usepackage{bm}

\usepackage[group-minimum-digits=3]{siunitx}

\usepackage{enumitem}

\usepackage{graphicx}
\usepackage{float}
\usepackage{hhline}
\usepackage{multirow}
\usepackage{booktabs}

\usepackage[authoryear]{natbib}

\begin{document}

\begin{center}
{\Large \bfseries How Many Customers Does a Retail Store Have?}\footnote{Preliminary results of the approach utilizing the least squares with different clustering and without the probabilistic background were presented in \cite{Sokol2018a}.}
\end{center}

\begin{center}
{\bfseries Ondřej Sokol} \\
University of Economics, Prague \\
Winston Churchill Square 4, 130 67 Prague 3, Czechia \\
\href{mailto:ondrej.sokol@vse.cz}{ondrej.sokol@vse.cz} \\
Corresponding Author
\end{center}

\begin{center}
{\bfseries Vladimír Holý} \\
University of Economics, Prague \\
Winston Churchill Square 4, 130 67 Prague 3, Czechia \\
\href{mailto:vladimir.holy@vse.cz}{vladimir.holy@vse.cz}
\end{center}

\begin{center}
{\itshape \today}
\end{center}

\noindent
\textbf{Abstract:}
The knowledge of the number of customers is the pillar of retail business analytics. In our setting, we assume that a portion of customers is monitored and easily counted due to the loyalty program while the rest is not monitored. The behavior of customers in both groups may significantly differ making the estimation of the number of unmonitored customers a non-trivial task. We identify shopping patterns of several customer segments which allows us to estimate the distribution of customers without the loyalty card using the maximum likelihood method. In a simulation study, we find that the proposed approach is quite precise even when the data sample is very small and its assumptions are violated to a certain degree. In an empirical study of a drugstore chain, we validate and illustrate the proposed approach in practice. The actual number of customers estimated by the proposed method is much higher than the number suggested by the naive estimate assuming the constant customer distribution. The proposed method can also be utilized to determine penetration of the loyalty program in the individual customer segments.
\\

\\

\\

\section{Introduction}
\label{sec:intro}

One of the most basic questions in retail business analytics is \emph{how many customers does a retail store actually have}.  In contractual situation (such as subscription services) or online environment (such as e-shops), it is quite easy to count unique customers. In retail brick-and-mortar stores, however, it is a much harder task as not all customers are directly monitored. In our setting, we assume that a portion of customers is monitored through the loyalty program while the rest of the customers can visit the store repeatedly without any possibility of identifying them. With absolute certainty, we can say that there is at least one customer without the loyalty card -- a single customer could theoretically purchase all unmonitored shopping baskets. We also know that there are at most as many customers without the loyalty card as unmonitored receipts -- each customer could visit the store only once. To be able to give a more specific estimate, we need to adopt further assumptions about the customer behavior. One may assume that the number of customers is proportional to the number of sales, i.e.\ customers with the loyalty card shop as often as customers without the loyalty card. In that case, the number of customers can be estimated simply by the number of sales divided by the average frequency of customer visits. However, this assumption is too strict and unrealistic as customers without the loyalty card are likely to visit the store less frequently. The question is how less frequently. The goal of our paper is to accurately estimate the number of unique customers under less restrictive and more realistic assumptions. We also estimate the distribution of unmonitored customers in specific customer segments.

The \emph{number of unique customers of a specified recency} is a valuable indicator of \emph{performance in attracting and retaining customers} as stressed by \cite{Bendle2015}. Such easily interpreted metric is of interest for both the management and shareholders. Furthermore, the number of unique customers in homogeneous customer segments with similar needs and purchase patterns brings an additional insight and consequently profit as the company can treat dissimilar customers differently. For example, \cite{Bendle2015} consider three tiers of customers based on their profitability for the company and suggest three respective strategies -- \emph{reward}, \emph{grow} and \emph{fire}. The analysis of customer segments instead of the identification of each individual customer is often enough as many companies are not able to work with customers on the individual level due to their large number (see e.g.\ \citealp{Berger2002}). The knowledge of the distribution of the user base is therefore a key aspect of strategic planning and finds its use in marketing departments involved in the retail decision-making process. It can be used to improve \emph{mass marketing communications}, especially by measuring the response to ads and promotional sales. The number of customers in certain segment can also be used as an input in \emph{prediction of sales} as each segment has different shopping behavior. All the above-mentioned applications play an important role in increasing the profit of the company. Accordingly, one of the main reasons for introducing the loyalty program to the store is to obtain information about the composition of customers and their purchasing behavior. The proposed approach allows the customer composition to be estimated for customers who are not willing to enter the loyalty program by utilizing only standardly collected data. As a result, it is quite sufficient that only a part of the customers enters the loyalty program. Our approach therefore allows to save costs of excessive promotion of the loyalty program while it estimates the complete customer composition. Finally, the proposed method can also be utilized in the field of \emph{customer-based corporate valuation} which ties the overall financial valuation of the company to the number of customers (see e.g.\ \citealp{Gupta2004a, McCarthy2018}). For a general overview of retail business analytics and its impact, see \cite{Lilien2013}, \cite{Germann2014}, \cite{Roberts2014}, \cite{Bradlow2017} and \cite{France2019}.

In the literature, counting of customers is most often approached in the context of \emph{active customers} that are monitored. \cite{Schmittlein1987} propose the \emph{Pareto/NBD Model} to determine the probability that a customer with a given pattern of transactions is still active. \cite{Schumacher2006} deals with \emph{faux-new customers} who seem to be new customers due to the absence of past transaction history but are actually regular customers. In contrast, we count customers that are never monitored.

In the proposed method, we consider that the customers who visited the store in a given period are split into the \emph{monitored sample} and the \emph{unmonitored sample}. The number of unique monitored customers is known as they can be identified using the loyalty card. On the other hand, the number of unique unmonitored customers is unknown as they cannot be linked to baskets purchased without the loyalty card. We propose the estimation of the number of unique unmonitored customers in a given period based solely on transaction data. Our method requires the following assumptions:
\begin{enumerate}[label=(\roman*)]
\item Customers are either completely monitored or unmonitored. All baskets purchased by monitored customers are linked to them.
\item Transactions are related to a single basket type and a single customer segment. Basket types and customer segments are same for the monitored and unmonitored samples. There is a positive number of customers in each segment in the monitored sample.
\item Basket types are observed for all transactions. Customer segments are observed only for transactions in the monitored sample.
\item Probabilities of individual basket types for a given customer segment in the unmonitored sample are same as the ones estimated from the monitored sample. These conditional probabilities are linearly independent.
\item Frequencies of visits by customers in individual segments are same for the monitored and unmonitored samples.
\end{enumerate}
In other words, we assume that the behavior of customers in a given customer segment is the same with and without the loyalty card while the distribution of customer segments itself can differ. For example, the customers with children (one of our customer segments) purchase similar types of baskets with similar shopping frequency whether they have the loyalty card or not. However, most of the customers with children do have the loyalty card and therefore they do not occur in the unmonitored sample as often. Under these assumptions, our procedure consists of the following steps:
\begin{enumerate}[label=(\roman*)]
\item We determine basket types based on the value and structure of the products in the basket. We also determine customer segments using purchase history of the customers with the loyalty card. For each customer segment, we have the average ratios of purchased basket types and the average shopping frequency.
\item We estimate the distribution of customer segments using the observed distribution of basket types not linked to loyalty cards. For this purpose, we utilize the maximum likelihood method.
\item We estimate the number of unique customers in a given period from the distribution of customer segments and average shopping frequencies.
\end{enumerate}
Using simulations, we show that the proposed estimator significantly outperforms the naive approach which assumes constant customer distribution. Furthermore the proposed estimator is quite precise even when the data sample is very small and our assumptions are violated to some extent. The proposed method is therefore very suitable in practice.

We apply the proposed approach to a Czech drugstore chain. First, we validate the method using only transactions linked to loyalty cards. For the purposes of this experiment, we redefine the monitored sample to contain only customers in the loyalty program with verified e-mail and the unmonitored sample to contain customers in the loyalty program without added e-mail. As we observe the actual number of unique customers in both samples, we can study the error of the proposed estimator in this case. We find that the mean absolute percentage error of the proposed estimator is on average $3.33$ percent in monthly periods and $4.38$ percent in quarterly periods. In contrast, it is  $13.33$ percent and $24.52$ percent respectively for the naive estimator assuming constant customer distribution. Second, we estimate the number of unique customers without the loyalty card and investigate their composition. We find that the number of customers estimated by the proposed method is on average $1.21$ higher in monthly periods and $1.31$ higher in quarterly periods than the number suggested by the naive estimator. As expected, we find that the distribution of customers without the loyalty card includes much more casual customers who visit the store rarely and purchase smaller baskets. On the other hand, we find that majority of regular customers are members of the loyalty program supporting the hypothesis that the loyalty card is very popular among regular customers and the loyalty program may be near the point of saturation.

The rest of the paper is structured as follows. In Section \ref{sec:meth}, we propose a procedure for determining the number and distribution of unmonitored customers. In Section \ref{sec:sim}, we examine the behavior of the proposed method using simulations. In Section \ref{sec:emp}, we illustrate the applicability of the proposed method in practice. In Section \ref{sec:disc}, we discuss various issues related to the proposed procedure. We conclude the paper in Section \ref{sec:con}.

\section{Methodology}
\label{sec:meth}

\subsection{Preliminary Clustering}
\label{sec:methClustering}

In the first step of our method, we cluster baskets to $n$ \emph{basket types} and customers to $m$ \emph{customer segments}. This step may vary from application to application. We do, however, impose the following property. For a customer segment $j$, $j=1,\ldots,m$, let $\hat{r}_{ij}^0$ denote the ratio of baskets belonging to a basket type $i$, $i=1,\ldots,n$. In the matrix notation, we have $\hat{r}^0 = (\hat{r}^0_{i,j})_{i=1,j=1}^{n,m}$. We then require that matrix $\hat{r}^0$ has linearly independent columns. This implies that the number of basket types must be greater than or equal to the number of customer segments, i.e.\ $n \geq m$. Another implication is that two customer segments cannot have the same ratios of basket types. The linear independence requirement is related to the identifiability of customer segments from the observed basket types. We further elaborate on the identifiability issue in Section \ref{sec:methSamples}.

Besides this "hard" rule, we also suggest the following "soft" rules which serve more as guidelines for the selection of suitable clustering techniques rather than strict limitations. Clustering of both baskets and customers should be based solely on the transaction history over a given time period. Besides the time frame, the time aspect should not be further utilized. Various characteristics of transactions related to purchased products such as the value, promotion, exclusivity, purpose, brand, diversity and category can be considered. Product categories can be either defined by an expert or found by yet another clustering analysis (see e.g.\ \citealp{Holy2017}). The customer segments should be selected in a way that the customers within the same segment have similar shopping frequency. We present the details of our clustering of basket types in Section \ref{sec:empBasketTypes} and clustering of customer segments in Section \ref{sec:empCustSegments}. Our clustering approach can easily be modified to suit specifics of other applications.

\subsection{Stochastic Framework}
\label{sec:methTheory}

We introduce our probabilistic framework and notation. Let $a$ denote the number of transactions. Further, let there be $n$ basket types and $m$ customer segments as in the previous section. Each transaction has a single basket type and a customer segment. Let $B_k \in \{ 1,\ldots,n\}$ be a random variable denoting the basket type of transaction $k$, $k=1,\ldots,a$, and $C_k \in \{ 1,\ldots,m\}$ a random variable denoting its customer segment. In this setting, we do not treat randomness from the perspective of individual customers, i.e.\ we do not model whether a specific customer decides to come to the store or not. Instead, we take the point of view of the cashier, i.e. we consider as given that a customer makes a purchase and we only wonder what kind of customer and purchase it is. In other words, random variables $B_k$ and $C_k$ are naturally conditioned on the fact that transaction $k$ occurs. This setting allows us to formulate a fairly simple model.

The random number of transactions with basket type $i$ is then $X_i = \sum_{k=1}^a \mathbb{I} \left\{ B_k = i \right\}$, $i = 1,\ldots,n$ while the random number of transactions with customer segment $j$ is $Y_j = \sum_{k=1}^a \mathbb{I} \left\{ C_k = j \right\}$, $j = 1,\ldots,m$, where $ \mathbb{I}(\cdot)$ denotes the indicator function. The random number of transactions with basket type $i$ and customer segment $j$ is $Z_{i,j} = \sum_{k=1}^a \mathbb{I} \left\{ B_k = i \wedge C_k = j \right\}$, $i = 1,\ldots,n$, $j = 1,\ldots,m$. Clearly, we have
\begin{equation}
\sum_{i=1}^n X_i = \sum_{j=1}^m Y_j = \sum_{i=1}^n \sum_{j=1}^m Z_{i,j} = a.
\end{equation}
In the matrix notation, we have $B = (B_1, \ldots, B_a)'$, $C = (C_1, \ldots, C_a)'$, $X = (X_1, \ldots, X_n)'$, $Y = (Y_1, \ldots, Y_m)'$ and $Z = (Z_{i,j})_{i=1,j=1}^{n,m}$. We denote $b = (b_1, \ldots, b_a)'$ the observed basket types and $c = (c_1, \ldots, c_a)'$ the observed customer segments. We further denote $x = (x_1, \ldots, x_n)'$ the observed numbers of transactions with given basket type, $y = (y_1, \ldots, y_m)'$ the observed numbers of transactions with given customer segment and $z = (z_{i,j})_{i=1,j=1}^{n,m}$ the observed numbers of transactions with given basket type and customer segment.

We assume that random variables $\{B_k:k=1,\ldots,a\}$ are independent and identically distributed with the probability of basket type $i$ denoted as $p_i = \mathrm{P}[B_k = i]$, $i = 1,\ldots,n$. Similarly, we assume that $\{C_k:k=1,\ldots,a\}$ are independent and identically distributed with the probability of customer segment $j$ denoted as $q_j = \mathrm{P}[C_k = j]$, $j = 1,\ldots,m$. The probability of basket type $i$ conditional on customer segment $j$ is denoted as $r_{i,j} = \mathrm{P}[B_k = i | C_k = j]$, $i = 1,\ldots,n$, $j = 1,\ldots,m$. In the matrix notation, we have $p = (p_1, \ldots, p_n)'$, $q = (q_1, \ldots, q_m)'$ and $r = (r_{i,j})_{i=1,j=1}^{n,m}$. Using the law of total probability, we have
\begin{equation}
\label{mainEq}
p_i = \mathrm{P} \left[ B_k = i \right] = \sum_{j=1}^m \mathrm{P} \left[ B_k = i | C_k = j \right] \mathrm{P} \left[ C_k = j \right] = \sum_{j=1}^m r_{i,j} q_j, \qquad i = 1,\ldots,n.
\end{equation}
In the matrix notation, we simply have $p = rq$. This is the key equation used in the proposed approach.

Finally, let $f_j$ denote the average frequency of visits by customers in segment $j$, $j=1,\ldots,m$. In the matrix notation, we have $f = (f_1, \ldots, f_m)'$. This is the only metric in our setting relating to individual customers.

\subsection{Observed Samples}
\label{sec:methSamples}

We consider that we observe two samples of transactions. In the \emph{monitored sample}, the transactions are linked to loyalty cards and we observe the basket types denoted as $b^{0}$, customer segments denoted as $c^0$ and average frequencies denoted as $f^{0}$. The \emph{unmonitored sample} contains transactions by customers without the loyalty card and we observe only basket types $b$. In terms of notation, we always use superscript 0 when referring to the monitored sample.

First, let us briefly investigate the case of the monitored sample. The probability of basket type $i$ can be simply estimated as $\hat{p}^0_i = x_i^0/a^0$, $i=1,\ldots,n$, the probability of customer segment $j$ as $\hat{q}^0_j = y_j^0/a^0$, $j=1,\ldots,m$ and the probability of basket type $i$ conditional on customer segment $j$ as $\hat{r}^0_{i,j} = z_{i,j}^0/y_j^0$, $i=1,\ldots,n$, $j=1,\ldots,m$.

Next, let us investigate the case of the unmonitored sample. We can estimate $p$ in the same way, i.e.\ $\hat{p}_i = x_i/a$, $i=1,\ldots,n$, but cannot estimate $q$ nor $r$. However, if we additionally assume that the matrix of conditional probabilities $r$ is known, we are able to estimate the customer segment distribution $q$. In our approach, we assume that the matrix of conditional probabilities in the unomnitored sample is the same as its estimate in the monitored sample, i.e.\ $r=\hat{r}^0$. In the rest of this section, we propose several estimators of $q$ under this setting. We utilize the equation $\hat{p} = r \hat{q}$. We remind that matrix $r$ has linearly independent columns as assumed in Section \ref{sec:methClustering}. The reason for this is that in the case of linearly dependent columns, the vector of customer probabilities $\hat{q}$ is not identifiable as the equation $\hat{p} = r \hat{q}$ with variable $\hat{q}$ has infinite number of solutions. If the matrix $r$ has linearly independent columns and is of square form, i.e.\ $n=m$, the estimation is quite simple as we can invert the matrix $r$ and obtain the straightforward estimator $\hat{q} = r^{-1} \hat{p}$. However, finding a meaningful basket types and customer segments satisfying the independence and square restriction can be quite difficult. Therefore, this case is theoretically ideal but unlikely to occur in practice. If the matrix $r$ has linearly independent columns and is tall, i.e.\ $n>m$, the estimation is more complex. In this case, the equation $\hat{p} = r \hat{q}$ can be inconsistent, i.e.\ without a solution for $\hat{q}$. However, we can find suitable $\hat{q}$ by minimizing an error between $\hat{p}$ and $r \hat{q}$. For this purpose, we utilize the maximum likelihood method. As this case is the most realistic, we devote to it the rest of the paper.

\subsection{Maximum Likelihood Estimation}
\label{sec:methLik}

Let us assume that we observe the vector of basket types $b$. The total number of transactions is denoted as $a$ while the vector of the numbers of transactions with given basket type is denoted as $x$. Let us further assume that the conditional probability matrix is $r=\hat{r}^0$ and has linearly independent columns with $n>m$. The \emph{maximum likelihood} method maximizes the likelihood function or, equivalently, the logarithm of the likelihood function. The logarithmic likelihood function is given by
\begin{equation}
\label{eq:lik}
\begin{aligned}
L(q) &= \ln \mathrm{P} \left[ B = b | Q = q \right] \\
&= \sum_{k = 1}^a \ln \mathrm{P} \left[ B_k = b_k | Q = q \right] \\
&= \sum_{i = 1}^n x_i \ln \mathrm{P} \left[ B_k = i | Q = q \right], \\
\end{aligned}
\end{equation}
where the second equality holds as $\{B_k\}$ are independent and the third equality holds as $\{B_k\}$ are identically distributed. Using the law of total probability, we have
\begin{equation}
\mathrm{P} \left[ B_k = i | Q = q \right] = \sum_{j=1}^m \mathrm{P} \left[ B_k = i | C_k = j, Q = q \right] \mathrm{P} \left[ C_k = j | Q = q \right].
\end{equation}
The logarithmic likelihood function is then
\begin{equation}
L(q) = \sum_{i = 1}^n x_i \ln \left( \sum_{j=1}^m r_{i,j} q_j \right).
\end{equation}
The estimates $\hat{q} = (\hat{q}_1, \ldots, \hat{q}_m)'$ are found by solving the nonlinear optimization problem
\begin{equation}
\begin{aligned}
\hat{q} \in \arg \max_{q} \ & \sum_{i = 1}^n x_i \ln \left( \sum_{j=1}^m r_{i,j} q_j \right) \\
\text{s.t.} \ & \sum_{j=1}^m q_j = 1, \\
& 0 \leq q_j \leq 1, \qquad j=1,\ldots,m.
\end{aligned}
\end{equation}
Note that this is equivalent to minimizing the Kullback--Leibler divergence from $rq$ to $\hat{p}$
\begin{equation}
KL(q) = \sum_{i = 1}^n \hat{p}_i \ln \left( \frac{\hat{p}_i}{\sum_{j=1}^m r_{i,j} q_j} \right).
\end{equation}
The \emph{Kullback--Leibler divergence} also known as \emph{relative entropy} measures how one probability distribution differs from a second probability distribution.

\subsection{Number of Unique Customers}
\label{sec:methNum}

Finally, we estimate the number of unique customers. Suppose that we have the vector $\hat{q}$ containing the estimated probabilities of the customer segments of a transaction. Let us further assume that the vector of frequencies is same for the observed and unobserved samples, i.e.\ $f = f^0$. If we observe $a$ transactions, the number of transactions with customer segment $j$ is estimated as $\hat{q}_j a$ for $j=1,\ldots,m$. The average number of unique customers in customer segment $j$ is estimated as $\hat{q}_j a / f_j$ for $j=1,\ldots,m$. The average total number of unique customers is estimated as
\begin{equation}
\hat{d} = \sum_{j=1}^m \frac{\hat{q}_j a}{f_j}.
\end{equation}
The probability of a customer belonging to customer segment $j$ is then estimated as
\begin{equation}
\hat{u}_j = \frac{\hat{q}_j a}{f_j \hat{d}}, \qquad j = 1,\ldots,m.
\end{equation}
Note that $\hat{q}$ denotes the estimated distribution of the customer segments of a \emph{transaction} while $\hat{u} = (\hat{u}_1, \ldots, \hat{u}_m)'$ denotes the estimated distribution of the customer segments of a \emph{customer}.

\section{Simulation Study}
\label{sec:sim}

\subsection{Design of the Simulation Study}

We investigate the finite-sample properties of the proposed approach using simulations. We utilize 9 simulation scenarios with various sample sizes and violated assumptions in order to analyze performance of the proposed method in comparison to the naive estimator which assumes constant customer distribution. In particular, we focus on the small size of the monitored and unmonitored samples, change in the conditional probability matrix and the frequency vector between the monitored and unmnonitored samples and linear dependence in the conditional probability matrix. For simplicity, we focus on the situation with $n=6$ basket types and $m=3$ customer segments. We consider both the basket types and the customer segments to be given in advance.

Let $\nu$ denote the number of simulations of a specific scenario. Furthermore, let $d_s$ denote the true number of customers and $\hat{d}_s$ its estimate in simulation $s$, $s=1,\ldots,\nu$. For the evaluation of the accuracy of the estimates, we use the \emph{mean absolute percentage error (MAPE)} given by
\begin{equation}
MAPE \left( \hat{d}_1,\ldots,\hat{d}_\nu \right) = 100\% \cdot \frac{1}{\nu} \sum_{s=1}^{\nu} \left| \frac{d_s - \hat{d}_s}{d_s} \right|.
\end{equation}
It is the average absolute difference between the true values and the estimated values divided by the true values. The estimates with lower MAPE are preferred. In some simulation scenarios, we have different values of the true number of customers. For example, the true number of unmonitored customers in scenario (ii) in Table \ref{tab:sim} is \num{500000} while it is \num{500} in scenario (iv). The naive estimates assuming the same distribution for customers in the monitored and unmonitored samples are \num{300000} in scenario (ii) and \num{300} in scenario (iv). The absolute error is \num{200000} and \num{200} respectively while the MAPE is 40 percent in both cases. The MAPE therefore allows us to compare differently scaled scenarios. This is the main motivation for utilizing the MAPE measure in our simulation study. \cite{McCarthy2006} and \cite{Fildes2007} survey forecasting practices in management and find that the MAPE is the most commonly used accuracy measure.

Next, we present the values of the parameters for the \emph{benchmark scenario} denoted as (ii) in Table \ref{tab:sim}. The customer distribution vector, the conditional probability matrix and the frequency vector for the monitored and unmonitored samples are respectively given by
\begin{equation}
\label{eq:simSetup}
q^{0} = 
\begin{pmatrix}
0.60 \\
0.20 \\
0.20 \\
\end{pmatrix}, \ 
q = 
\begin{pmatrix}
0.20 \\
0.20 \\
0.60 \\
\end{pmatrix}, \ 
r^0 = 
\begin{pmatrix}
0.40 & 0.10 & 0.10 \\
0.20 & 0.10 & 0.10 \\
0.10 & 0.40 & 0.10 \\
0.10 & 0.20 & 0.10 \\
0.10 & 0.10 & 0.40 \\
0.10 & 0.10 & 0.20 \\
\end{pmatrix}, \ 
r = r^0, \ 
f^{0} = 
\begin{pmatrix}
6.00 \\
3.00 \\
1.50 \\
\end{pmatrix}, \ 
f = f^0.
\end{equation}
In the monitored sample, most customers shop with relatively high frequency while in the unmonitored sample, most customers shop with much lower frequency. Each customer segment has its own two dominant basket types. Similar behavior is observed in the empirical study in Section \ref{sec:emp}. Furthermore, the number of observations is $a^0 = \num{1000000}$ for the monitored sample and $a = \num{1000000}$ for the unmonitored sample. Such values of the customer distribution vector, the frequency vector and the number of observations result in the number of customers $d^0 = \num{300000}$ in the monitored sample and $d = \num{500000}$ in the unmonitored sample. Note that the customer distribution vector changes in the monitored and unmonitored samples while the conditional probability matrix and the frequency vector remain the same. This is consistent with our assumptions. If the customer distribution vector does not change, the naive estimator is superior and there is no need for the proposed approach as we demonstrate in scenario (i) in Table \ref{tab:sim}. This is, however, unrealistic in practice.

We base all the other scenarios on the benchmark scenario (ii) with some modifications. Scenarios (iii)--(v) are designed to assess the effects of sample sizes while scenarios (vi)--(ix) are designed to investigate the effects of assumption violations.

\begin{table}
\begin{center}
\begin{tabular}{llrrrrrr}
\toprule
\multicolumn{2}{c}{Simulation Scenario} & \multicolumn{3}{c}{Naive Method} & \multicolumn{3}{c}{Proposed Method} \\
No. & Description & M & SD & 95\% & M & SD & 95\% \\
\midrule
  (i) & No change in $q$ & 0.00 & 0.00 & 0.00 & 0.19 & 0.14 & 0.46 \\ 
  (ii) & Benchmark scenario & 40.00 & 0.00 & 40.00 & 0.18 & 0.14 & 0.45 \\ 
  (iii) & Small $a^0 = 1000$ & 40.00 & 0.00 & 40.00 & 5.18 & 4.01 & 12.66 \\ 
  (iv) & Small $a = 1000$ & 40.00 & 0.00 & 40.00 & 2.61 & 1.97 & 6.41 \\ 
  (v) & Small $a^0 = a = 1000$ & 40.00 & 0.00 & 40.00 & 5.80 & 4.47 & 14.14 \\ 
  (vi) & Change in $r$ & 40.00 & 0.00 & 40.00 & 5.12 & 3.85 & 12.50 \\ 
  (vii) & Change in $f$ & 40.95 & 10.97 & 58.97 & 14.76 & 10.92 & 35.69 \\ 
  (viii) & Change in $r$ and $f$ & 40.95 & 10.97 & 58.97 & 15.62 & 11.56 & 37.85 \\ 
  (ix) & Linear dependence in $r$ & 40.00 & 0.00 & 40.00 & 30.13 & 15.96 & 50.06 \\ 
\bottomrule
\end{tabular}
\caption{Mean values (M) of absolute percentage errors with standard deviations (SD) and 95\%-quantiles (95\%) of the naive and proposed estimates for 9 scenarios based on \num{300000} simulations.}
\label{tab:sim}
\end{center}
\end{table}

\subsection{Sample Size}
\label{sec:simSize}

\begin{figure}
\begin{center}
\includegraphics[width=130mm]{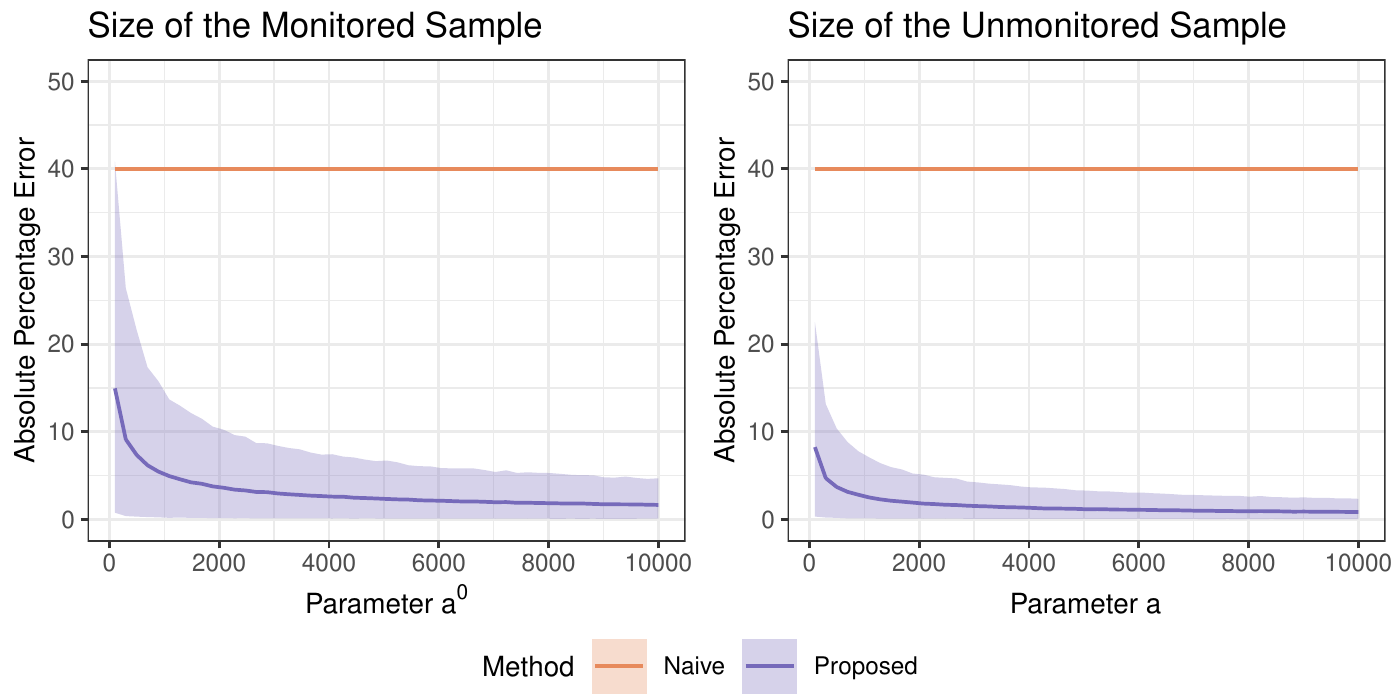} 
\caption{Mean absolute percentage errors with 95\% confidence intervals of the naive and proposed estimates based on \num{5000} simulations with various sample sizes.}
\label{fig:sample}
\end{center}
\end{figure}

We consider various sizes of the monitored and unmonitored samples. Specifically, the size of the monitored sample $a^0$ ranges from \num{100} to \num{10000} in the left plot of Figure \ref{fig:sample} and is set to \num{1000} in scenarios (iii) and (v) in Table \ref{tab:sim}. Similarly, the size of the unmonitored sample $a$ ranges from \num{100} to \num{10000} in the right plot of Figure \ref{fig:sample} and is set to \num{1000} in scenarios (iv) and (v) in Table \ref{tab:sim}. In the extreme case of 100 observations, the probability that some customer segment is not present in the monitored data sample is $4 \cdot 10^{-10}$ for the customer distribution vector and the frequency vector given by \eqref{eq:simSetup}. In our simulations, all customer segments are always present. As the customer segmentation is build on the monitored sample in real applications, the situation with missing customer segments would not make any sense.

As we can see in Figure \ref{fig:ass}, the proposed estimator performs quite well even for very small monitored and unmonitored data samples. The size of the monitored sample is more important in comparison to the size of the unmonitored sample as the monitored sample is used to estimate the conditional probability matrix and the frequency vector. Overall, the sample size is not really an issue as retail stores have typically large amount of data available.

\subsection{Violation of Assumptions}
\label{sec:simAss}

\begin{figure}
\begin{center}
\includegraphics[width=130mm]{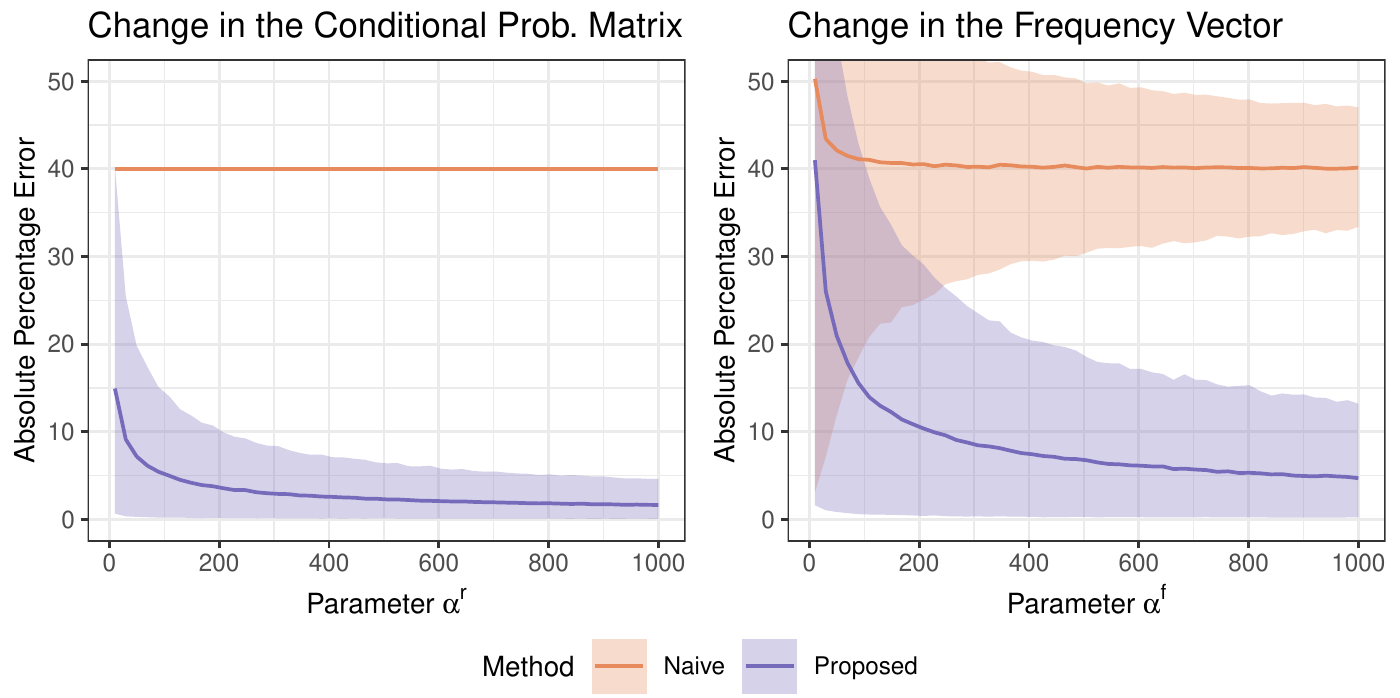} 
\caption{Mean absolute percentage errors with 95\% confidence intervals of the naive and proposed estimates based on \num{5000} simulations with violated assumptions.}
\label{fig:ass}
\end{center}
\end{figure}

First, we examine the change in the conditional probability matrix in the unmonitored sample. Let $r^0_j$ denote the $j$-th column of matrix $r^0$ and $r_j$ denote the $j$-th column of matrix $r$. For the purposes of the simulation study, we consider the columns $r_j$ to be random vectors following the Dirichlet distribution with the concentration parameter equal to $\alpha^r r^0_j$. Parameter $\alpha^r$ controls to what degree is the assumption $r = r^0$ violated. Small values indicate significant deviation from $r^0$ while for $\alpha^r \to \infty$, random vector $r$ converges to deterministic value $r^0$. The value of $\alpha^r$ ranges from \num{10} to \num{1000} in the left plot of Figure \ref{fig:ass} and is set to \num{100} in scenarios (vi) and (viii) in Table \ref{tab:sim}. This parameter is, however, a bit cumbersome to interpret. From the Bayesian point of view, $\alpha^r r^0_j$ is the increment of the concentration parameter of the Dirichlet distribution after $\alpha^r$ random variables distributed according to $r^0_j$ are observed. The average absolute difference between the elements of $r^0$ and $r$ is 0.08 for $\alpha^r=\num{10}$, 0.03 for $\alpha^r=\num{100}$ and 0.01 for $\alpha^r=\num{1000}$.

Second, we examine the change in the frequency vector in the unmonitored sample. In a similar fashion, we consider the frequency vector $f$ to be random and to follow the Dirichlet distribution. In this case, however, the sum of the random vector is set to the sum of $f^0$ instead of 1. Again, we control to what degree is the assumption $f = f^0$ violated using parameter $\alpha^f$. The value of $\alpha^f$ ranges from \num{10} to \num{1000} in the right plot of Figure \ref{fig:ass} and is set to \num{100} in scenarios (vii) and (viii) in Table \ref{tab:sim}. The average absolute difference between the elements of $f^0$ and $f$ is 1.10 for $\alpha^f=\num{10}$, 0.36 for $\alpha^f=\num{100}$ and 0.11 for $\alpha^f=\num{1000}$. Note that a change in the frequency vector $f$ causes a change in the number of unmonitored customers $d$, which is also a random variable in this case.

Finally, we examine linear dependence in the conditional probability matrix. In scenario (ix) in Table \ref{tab:sim}, we set the last column in matrices $r^0$ and $r$ to be the average of the other columns.

The proposed method is more sensitive to changes in the frequency vector $f$ than to changes in the conditional probability matrix $r$ as we can clearly see in Figure \ref{fig:ass}. However, even for value $\alpha^f = 100$ with average absolute difference 0.36 in scenario (vii), the proposed estimator significantly outperforms the naive estimator. As shown in scenario (ix), linear dependence in $r$ can cause serious problems. The error caused by linear dependence is affected by the frequency vector $f$. For example, if there are two identical columns in $r$ and their corresponding frequencies are the same, there is no error. If the frequencies, however, significantly differ (as they do in our case), the error can be substantial. Luckily, linear dependence in $r$ can be avoided by selecting suitable segmentation using the monitored sample.

\section{Empirical Study}
\label{sec:emp}

\subsection{Background and Motivation}
\label{sec:empMotiv} 

In the empirical study, we examine data from a Czech drugstore retail chain. Like most large retailers, it has an established loyalty program. The loyalty program is a valuable source of data for the company as it enables not only better insight into customer behavior, but also crucial information about the structure of the client base. The purchase history of individual customers provides valuable resources for various analyses. For this reason, the company is trying to promote its program and attract new members. The company offers free membership in its loyalty program to all customers. For its members, the program brings a lot of benefits such as extra discounts and gifts for buying specific products. Joining the loyalty program is simple and fast -- it can be done quickly during checkout. Thanks to these perks, the loyalty program is very popular among the customers. Majority of receipts are therefore linked to a specific customer through the loyalty card.  Among the management, it is believed that the loyalty program may be already saturated among valuable customers, e.g.\ the vast majority of valuable customers who are willing to join the loyalty program are already members.
 
Unfortunately, the company does not have any reliable estimate of the composition of non-member customers based on hard data. We remedy this by achieving the following two objectives:
\begin{enumerate}[label=(\roman*)]
\item To estimate the number of unique customers among non-members.
\item To estimate the distribution of non-members grouped by customer segments. 
\end{enumerate}

\subsection{Transaction Data}
\label{sec:empData} 

Our dataset consists of individual purchase data of one of the retail chains in the Czech drugstore market. The retail chain sells over 10 thousand products which are divided into 55 categories. This categorization is done by an expert's opinion based on the product properties and purpose. Each product is also assigned to one of the three price-levels -- \emph{low-end}, \emph{standard} and \emph{high-end}. We use the data from the table of all transactions during the years 2017 and 2018. The dataset includes every single product sold, identification of the basket and in some cases link to a specific customer through the loyalty program.

The monitored sample consists of the purchases by the loyalty club members and the unmonitored sample consists of purchases without the loyalty card. We remove monitored customers with extremely high frequency (more than 15 visits per month) from the dataset as they are a special type of customers and we believe that no customer without the loyalty card exhibits this behavior. The cleaned dataset consists of over 50 million receipts, with 69 percent linked to a specific customer through the loyalty card.

\subsection{Basket Types}
\label{sec:empBasketTypes}

In order to get meaningful customer segmentation, we use a two phase method involving segmentation of both baskets and customers. The basic requirement is that customer segments have different conditional probabilities of basket types. We propose a simple approach consisting of clustering of baskets and customers using the k-means method over four dimensions. The k-means clustering has indisputable advantage in the simple interpretation using the centers of clusters. Clustering dimensions of both baskets and customers are selected with respect to predicted customer groups and may vary by store type. For example, the share of children products in basket clustering is included as we want to point out baskets focused on children. The parents of small children in drugstore are usually frequent customers often focused on price and sales promotions. Their shopping behavior is different and such customers may require special attention from the marketing point of view.

First, we determine basket types using k-means over transaction data aggregated by basket. For each basket we compute its \emph{value}, the share of \emph{premium products}, the share of \textit{products for children} and the \emph{diversity} of the basket. The value of the basket is standardized by the 95\% quantile. For each basket, we therefore get a number between 0 and 1 with 1 assigned to baskets with the value of the 95\% quantile and higher. The share of premium products is the value of premium products divided by the value of the whole basket. Similarly, we define the share of products for small children such as diapers, wipes and baby food. The diversity of the basket is determined by the number of products standardized by the 90\% quantile as we want to distinguish whether the basket contains multiple categories or just one or two. Each indicator is further standardized to zero mean and unit variance.

The combination of these four dimensions then allows us to distinguish different types of baskets. The optimal number of basket types is determined using the Davies--Bouldin statistic which is minimized in the case of 13 centers. However, the proposed method is robust to the choice of the number of centers as the differences in the results are minimal for the number of centers ranging from 10 to 15. Note that we find the basket types for each period separately. However, the approach converges to very similar clusters regardless of the chosen period.

\subsection{Customer Segments}
\label{sec:empCustSegments} 

Similarly to the first phase, we use four dimensions for clustering of customers. Instead of a single basket, however, we aggregate the four indicators over all baskets purchased by a customer in a given period. Specifically, we have the total \emph{value} of the customer, the share of \emph{premium products}, the share of \textit{products for children} and the average \emph{diversity}. Again, we standardize the indicators to zero mean and unit variance.

We use the k-means method to determine customer segments. Using the results of the empirical validation in Section \ref{sec:empValidation}, we find that the optimal number of customer segments is 5. Also, this is the number of clusters that minimizes the Davies--Bouldin index. With higher number of customer segments, k-means tends to find segments with similar probability of basket types but different frequencies. This results in various inaccuracies in estimates. In practice, the rows of matrix $r$ should be checked for similarity and in the case of high level of similarity, the suitability of the dimensions used for segmentation should be reconsidered. As in the case of basket types, we find the customer segments separately for each period. The centers are quite stable in time and their interpretation is the same for all periods.

The descriptions of the customer segments with their shopping frequencies are presented in Table \ref{tab:customerSegments}. The largest group among customers with the loyalty card are two types of general customers distinguished by the share of premium products -- the \emph{general premium} and \emph{general budget} customers. Those are the archetypes of general customers exhibiting standard behavior and moderate shopping frequency. The \emph{supplier} segment is formed by the customers who visit the store rarely but their value is high and their typical basket value is very high. The most valuable customers, however, belong to the \emph{loyal customer} segment with very high frequency, moderate average diversity and very high value. While segments of loyal customers and suppliers make just around one quarter of all members of loyalty program, they make over 55 percent of revenue. The customers who buy products for small \emph{children} are also valuable customers as their frequency is very high and their revenue moderate. However, margins of their most frequently purchased products are below standard. For this reason, we distinguish them as a special segment as the goal is not only to maximize their revenue but mainly redirect them to the products with higher margins. 

\begin{table}
\begin{center}
\begin{tabular}{llllll}
\toprule
Segment         & Value     & Premium   & Children  & Diversity & Frequency \\
\midrule
General Budget  & Very Low  & Very Low  & Low       & Low       & Moderate  \\
General Premium & Low       & High      & Low       & Low       & Moderate  \\
Loyal Customer  & Very High & Moderate  & Low       & Moderate  & Very High \\
Supplier        & High      & Moderate  & Low       & High      & Low       \\
With Children   & Moderate  & Low       & Very High & Low       & High      \\
\bottomrule
\end{tabular}
\caption{Description of the customers segments.}
\label{tab:customerSegments}
\end{center}
\end{table}

\subsection{Empirical Validation}
\label{sec:empValidation}

In order to validate the method using empirical data, we take the transaction data of customers with the loyalty card and divide it into two samples. In this section only, the monitored sample includes the customers who added e-mail address to their loyalty account and the unmonitored sample contains the rest of the customers in the loyalty program. Therefore, the customers in the monitored sample must provide additional data, while the customers in the unmonitored sample do not have to take any further action aside of simple joining the loyalty program. This partition is selected this way to emulate the differences between members and non-members of the loyalty program. 

\begin{figure}
\begin{center}
\includegraphics[width=130mm]{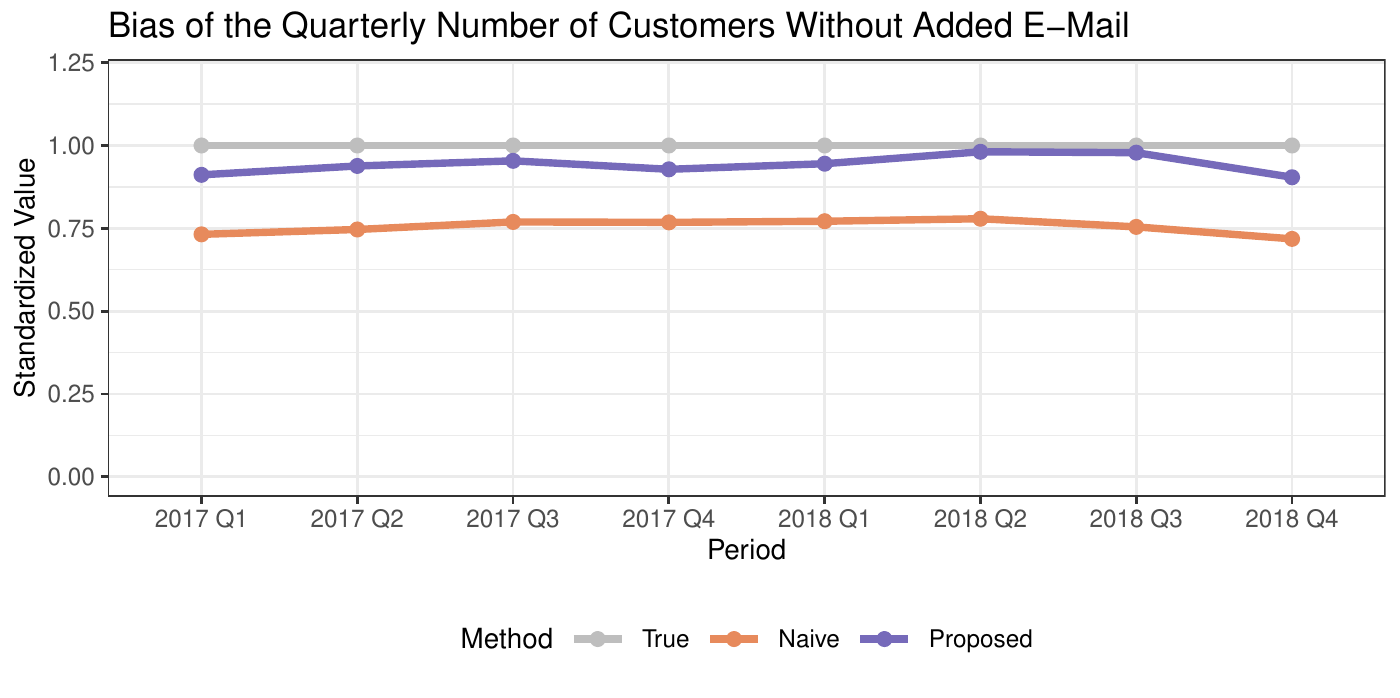}
\caption{Comparison of the proposed and naive estimates of the quarterly number of customers in the loyalty program without added e-mail standardized by the true values from 2017 to 2018.}
\label{fig:valTime}
\end{center}
\end{figure}

Table \ref{tab:val} highlights the performance of the method for different time frames using data from 2017 and 2018. As in the simulation study, we utilize the mean absolute percentage error based on the scaled difference between the estimated values and the true values. We compare the proposed method to the naive approach and find that the proposed method is superior in all considered time frames. In most cases, the number of customers is underestimated by both methods as illustrated in Figure \ref{fig:valTime} for the quarterly periods, although the proposed method is nevertheless quite precise. The accuracy of both estimators decreases as the time period increases. Decreasing accuracy seems to be related to the customer segmentation. The difference in frequencies in the monitored and unmonitored samples is significantly smaller for weekly data than for quarterly and especially annual data as majority of customers tend to visit the store only once per week.

\begin{figure}
\begin{center}
\includegraphics[width=130mm]{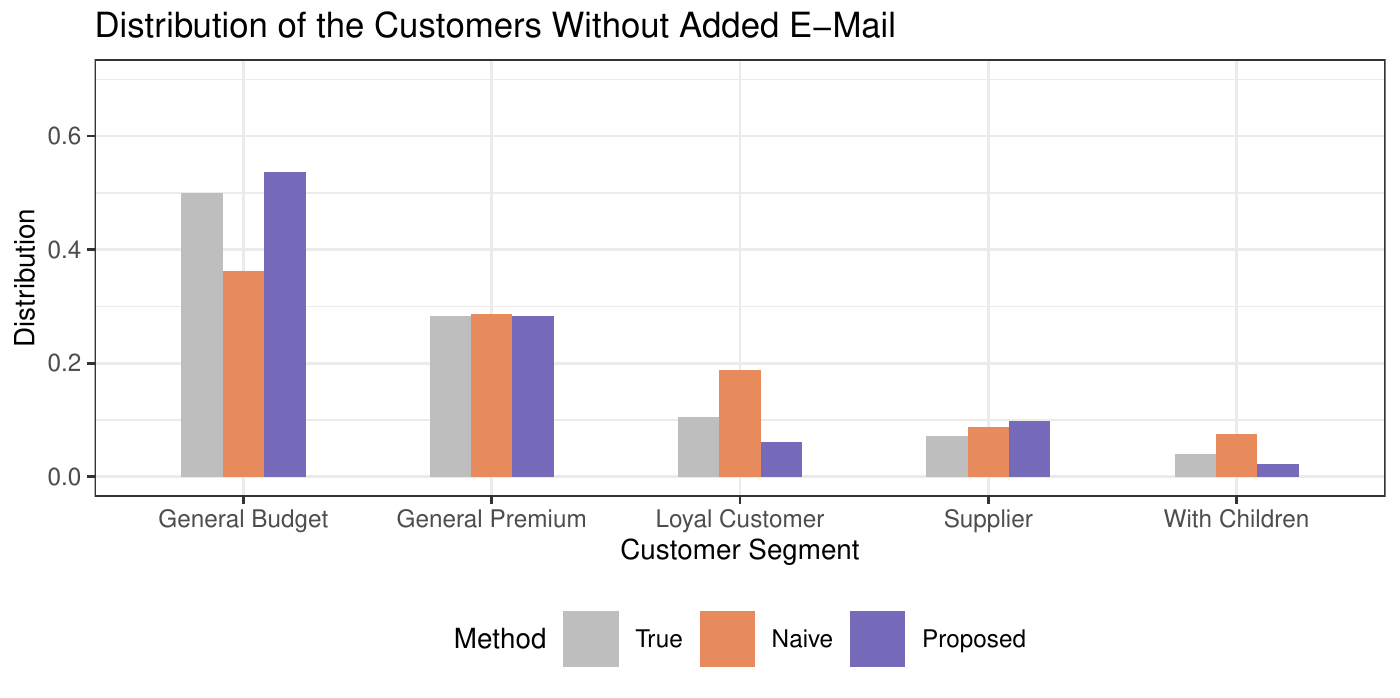}
\caption{Comparison of the proposed and naive estimates with the true values of the distribution of customers in the loyalty program without added e-mail in the second quarter of 2018.}
\label{fig:valDist}
\end{center}
\end{figure}

From our point of view, the most interesting time frame is the quarter, because there are already significant differences between the frequencies of individual segments. The comparison of estimated distribution of segments by the proposed and naive method in the second quarter of 2018 is shown in \ref{fig:valDist}. The estimates are also compared to the true values which are known in this validation. It is apparent that the proposed method has significantly higher accuracy than the naive estimation. The highest differences between methods relate to the segment of general budget customers. While the naive method significantly underestimates the number, the proposed method estimate is much closer to the true value, even though the estimate is slightly overstated. The opposite, in terms of the error direction, happens in the segment of loyal customers and customers with children (although at a much smaller scale).

\begin{table}
\begin{center}
\begin{tabular}{lrrrrrrr}
\toprule
\multicolumn{2}{c}{Data Sample} & \multicolumn{3}{c}{Naive Method} & \multicolumn{3}{c}{Proposed Method} \\
Period &  Obs. & M & SD & WC & M & SD & WC \\
\midrule
Week & 101 & 2.99 & 0.74 & 5.35 & 1.61 & 0.70  & 3.57 \\ 
Month & 24 & 13.33 & 2.04 & 16.69 & 3.33 & 1.56 & 8.20 \\ 
Quarter & 8 &24.52 & 2.09 & 28.19 & 4.38 & 2.00 & 9.59 \\ 
Year & 2 & 34.90 & 2.39 & 36.59 & 16.79 & 1.88 & 18.13 \\ 
\bottomrule
\end{tabular}
\caption{Mean values (M) of absolute percentage errors with standard deviations (SD) and the worst case (WC) of the naive and proposed estimates based on various periods from 2017 to 2018.}
\label{tab:val}
\end{center}
\end{table}

\subsection{Distribution of Non-Member Customers}
\label{sec:empDistribution}

\begin{figure}
\begin{center} 
\includegraphics[width=130mm]{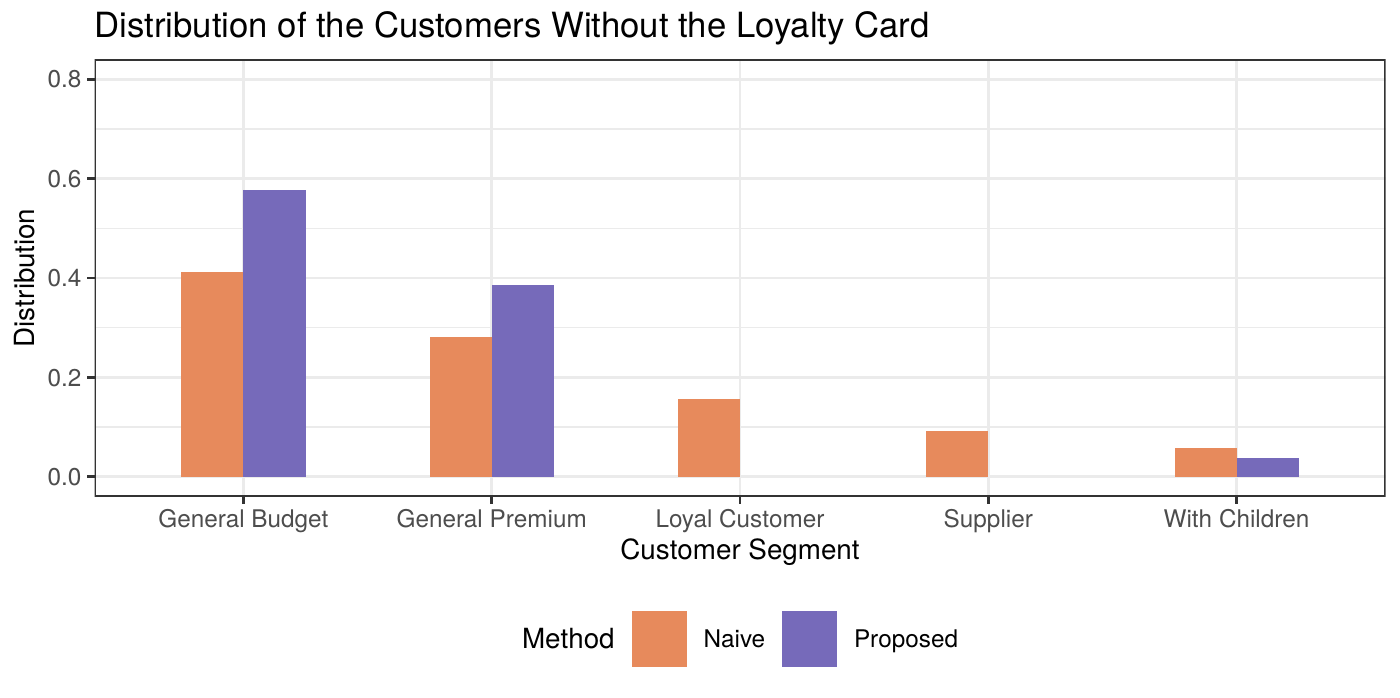} 
\caption{Comparison of the proposed and naive estimates of the distribution of customers without the loyalty card in the second quarter of 2018.}
\label{fig:bar}
\end{center}
\end{figure}

Now we come to the main goal -- the estimation of non-member customer composition during a given period. We focus on the second quarter of 2018 as we have an expert estimate available for this quarter. As we stated, it is reasonable to assume that the distribution of members and non-members of the loyalty program is substantially different and our case study of the drugstore retail chain supports this notion. The distribution of non-members estimated by the proposed and naive methods is shown in Figure \ref{fig:bar}. Note that the distribution given by the naive approach is identical to the distribution of members.

The results support the hypothesis that the loyalty program is saturated among the most valuable customers. Those customers are already members of the loyalty program in the loyal customer and suppliers segments. Both segments are virtually absent in the non-member group. The distribution of other customer segments is also significantly different from the distribution of members. While customers in general budget and general premium segments together form just 69 percent of the loyalty program members, they make almost 96 percent in the non-member sample. Customers with children make 6 percent of customers with the loyalty card and 4 percent of customers without the loyalty card. Using our estimates also allows us to compute penetration of each segment. The general budget customers are members of the loyalty program in 53 percent, general premium customers in 54 percents and customers with children in 71 percents.

Due to the estimated composition of non-member customers, the company can better plan marketing events to promote sales. For example, marketing activities that have proven successful with supplier and loyalty customer segments will have minimal success if targeted at non-member customers as those segments are not present in that sample. 
On the other hand, information about a relatively large proportion of clients who prefer premium products is essential because it undermines the idea that non-members customers, who visit the store rather rarely, are not willing to spend money on quality brand products. This is a kind of information that is not obtainable from the standard analysis of the sales data as it is not clear whether these purchases are made by customers who visit the store frequently or rarely. Our approach is able to clarify that. 

The differences in customer segment distributions between members and non-members of the loyalty program have a major influence on the estimation of the total number of unique non-member customers. The number estimated by the naive approach is 24 percent lower than the number estimated by the proposed method. In particular, the naive approach estimates \num{748196} non-member customers while the proposed method estimates \num{980568} customers. The proposed estimate is in line with the estimate of an expert, which we have available for this quarter only.
 
\section{Discussion}
\label{sec:disc}

\subsection{Time Aspect}
\label{sec:discTime}

Unlike other approaches such as the Pareto/NBD Model of \cite{Schmittlein1987}, we do not incorporate any sort of timing into the model, beyond the count of transactions over the observation period. Although the information about the timing of purchases may seem essential, it undermines the simplicity of the model. Adding a time component related to individual customers would ultimately result in pairing the individual baskets with individual customers. The model in this case would be significantly more complex and would require a lot of additional data and assumptions. This goes hand in hand with the emergence of new sources of estimation errors. For these reasons, we keep our proposed model as simple as possible. Nevertheless, adding the timing of purchases could prove to be an interesting topic for future research.

\subsection{One-and-Done Customers}
\label{sec:discDone}

It is common in retail that there are many \textit{one-and-done} customers who make just one purchase and never come back to the store in the future (see e.g. \citealp{Fader2005a}). Such customers are a bit hard to identify using the proposed approach as it is likely they do not join the loyalty program. However, there is a portion of customers with the loyalty card who visit the store rather sporadically -- often just once in the observed period. In our study, these customers are typically assigned to the general budget and general premium segments. We believe that their behavior is similar to the true one-and-done customers, especially if we focus on longer periods such as quarters. We therefore have homogeneous customer segments that are present in both monitored and unmonitored samples that containt one-and-done customers. In general, however, the proposed method cannot deal with customer segments that occur only in the unmonitored sample.

\subsection{Unused Loyalty Cards}
\label{sec:discLoyalty}

One of our assumptions is that monitored customers always use the loyalty card with every purchase. In our empirical study, customers are quite motivated to use the loyalty card for the following two reasons. First, the use of the loyalty card is convenient as customers can identify themselves either by the physical card or the mobile application. Second, prices are lower in the vast majority of baskets for the members of the loyalty program. Therefore, we believe that this assumption is justified in our case.

For other retail chains, however, this assumption may not be met. In this case, it would be necessary to estimate what portion of baskets purchased by customers in the loyalty club is made without identification. For this purpose, a customer survey or an expert estimate could be utilized. The rigorous estimation procedure is, however, beyond the scope of the paper. When the portion of purchases without identification is estimated, the number of unique customers can be straightforwardly adjusted.

\subsection{Identification by Payment Cards}
\label{sec:discPay}

Using payment card numbers to identify customers is a problem due to security concerns. While the transaction data are usually available for the analytic department, the payment card numbers are often unobtainable. Furthermore, the payment method of individual customers can change with each transaction making it impossible to reliably link customers and purchases. Even more, the choice between the payment by card and cash could depend on the value of the purchase. In the case of our dataset and many others, the cash payment is still very popular. Note that this is different from our assumption that monitored customers always use the loyalty card when making a purchase as there are no advantages of using the same payment method.

\subsection{Least Squares Estimation}
\label{sec:discSquares}

As an alternative to the maximum likelihood method presented in Section \ref{sec:methLik}, one may utilize the \emph{least squares} method. It is based on the minimization of the squared error given by
\begin{equation}
SE(q) = \sum_{i = 1}^n \left( \hat{p}_i - \sum_{j=1}^m r_{i,j} q_j \right)^2.
\end{equation}
In the simulation study, we find that the maximum likelihood estimator slightly outperforms the least squares estimator in terms of lower mean and standard deviation of the absolute percentage error in all nine scenarios. The difference between the two methods is, however, very small. In the empirical study, both methods also give very similar results. In conclusion, we recommend to use the maximum likelihood estimator based on our simulation study but the least squares estimator is nevertheless a very decent alternative.

\subsection{Bayesian Inference}
\label{sec:discVayes}

If we have prior knowledge about the distribution of customers, e.g.~an expert opinion or a previous study, we can adopt the \emph{Bayesian approach}. For a review of the use of Bayesian methods in marketing, see \cite{Rossi2003}. In Bayesian statistics, the \emph{maximum a posteriori} method maximizes the posterior probability density. Equivalently, we can maximize function
\begin{equation}
AP(q) = \ln \mathrm{P} \left[ B = b | Q = q \right] + \ln g_{Q}(q),
\end{equation}
where $g_{Q}(q)$ is the prior probability density of random parameter $Q$. Note that the first term is the logarithmic likelihood function \eqref{eq:lik}. The Dirichlet distribution is the most common choice for the prior distribution of a categorical variable as it is its conjugate prior distribution. In the case of the Dirichlet prior, we maximize function
\begin{equation}
AP(q) = \sum_{i = 1}^n x_i \ln \left( \sum_{j=1}^m r_{i,j} q_j \right) + \sum_{j=1}^m (\gamma_j - 1 ) \ln q_j - \ln B(\gamma),
\end{equation}
where $\gamma = (\gamma_1, \ldots, \gamma_m)'$ is the vector of concentration parameters capturing our prior knowledge and $B(\cdot)$ denotes the multivariate beta function. This approach can be viewed as a regularized version of the maximum likelihood method with apriori knowledge of distribution of parameters $Q$.

\subsection{Computational Issues}
\label{sec:discComp}

When operating with large databases of transactions, computational performance is always a concern. The most demanding step of the proposed method in terms of computing speed and memory usage is the preliminary clustering. The details, of course, depend on the selected clustering algorithm. In the next step, the number of transactions for the individual basket types and customer segments together with the average frequencies for the individual customer segments are computed by a simple aggregation procedure. The final steps of the estimation of the customer distribution and the number of customers work with data of very low dimension and are not computationally demanding at all.

For the maximum likelihood estimation, a variety of non-linear optimization algorithms can be utilized. In our simulation and empirical study, we adopt the \emph{improved stochastic ranking evolution strategy} algorithm of \cite{Runarsson2005}.

\section{Conclusion}
\label{sec:con}

We propose a method for estimation of the number of unique customers of a specified recency using retail transaction data. The method also estimates the number of unique customers in each customer segment. We investigate the finite-sample performance of the proposed method using synthetic data in the simulation study. The method performs quite well even when the data sample is very small and its assumptions are violated to some extent. Therefore, we expect the method to give reliable results in real applications.

In the empirical part, we analyze retail transaction data from a Czech drugstore retail chain. First, we validate the method using fully observed transactions. Second, we estimate the number of unique customers and quantify the initial proposition that the loyalty program is very popular among the most valuable customers. This is something which has been done only by an expert opinion before and therefore the estimates were contaminated by high uncertainty.

The reliable quantification of the customer composition is the main contribution of the paper. This is also the most desired result of the loyalty program. The number of unique customers is a valuable indicator of performance in attracting and retaining customers and can be further utilized in mass marketing communications, prediction of sales and customer-based corporate valuation.

\section*{Acknowledgements}
\label{sec:acknow}

We would like to thank Michal Černý, Miroslav Rada and anonymous reviewers for their useful comments. We would also like to thank organizers and participants of the 7th International Conference on Management (Nový Smokovec, September 26--29, 2018) and 7th International Business and Management Sciences Congress (Istanbul, March 14--15, 2019) for fruitful discussions.

\section*{Funding}
\label{sec:fund}

This work was supported by the Internal Grant Agency of the University of Economics, Prague under Grant F4/58/2017.


\end{document}